\documentclass[apj]{emulateapj}

\begin{document}

\shorttitle{X-rays and $\gamma$-rays from PSR J1723--2837}
\shortauthors{Bogdanov et al.}

\title{X-ray and $\gamma$-ray Studies of the Millisecond Pulsar and Possible \\ X-ray Binary/Radio Pulsar Transition Object PSR J1723--2837}

\author{Slavko Bogdanov\altaffilmark{1},  Paolo Esposito\altaffilmark{2}, Fronefield Crawford III\altaffilmark{3},\\  Andrea Possenti\altaffilmark{4}, Maura A.~McLaughlin\altaffilmark{5}, Paulo Freire\altaffilmark{6}} 

\altaffiltext{1}{Columbia Astrophysics Laboratory, Columbia University, 550 West 120th Street, New York, NY 10027, USA; slavko@astro.columbia.edu}

\altaffiltext{2}{INAF-IASF Milano, via E. Bassini 15, I-20133, Milano, Italy}

\altaffiltext{3}{Department of Physics and Astronomy, Franklin and Marshall College, P.O.~Box 3003, Lancaster, PA 17604, USA}

\altaffiltext{4}{INAF-Osservatorio Astronomico di Cagliari, Loc.~Poggio dei Pini, Strada 54, I-09012 Capoterra (CA), Italy}

\altaffiltext{5}{Department of Physics and Astronomy, West Virginia University, 210E Hodges Hall, Morgantown, WV 26506, USA}

\altaffiltext{6}{Max-Planck-Institut f\"ur Radioastronomie, D-53121 Bonn, Germany}

\begin{abstract}  
  We present X-ray observations of the ``redback'' eclipsing radio
  millisecond pulsar and candidate radio pulsar/X-ray binary
  transition object PSR J1723--2837. The X-ray emission from the
  system is predominantly non-thermal and exhibits pronounced
  variability as a function of orbital phase, with a factor of $\sim$2
  reduction in brightness around superior conjunction.  Such temporal
  behavior appears to be a defining characteristic of this variety of
  peculiar millisecond pulsar binaries and is likely caused by a
  partial geometric occultation by the main-sequence-like companion of
  a shock within the binary.  There is no indication of diffuse X-ray
  emission from a bow shock or pulsar wind nebula associated with the
  pulsar.  We also report on a search for point source emission and
  $\gamma$-ray pulsations in \textit{Fermi} Large Area Telescope data
  using a likelihood analysis and photon probability weighting.
  Although PSR J1723--2837 is consistent with being a $\gamma$-ray
  point source, due to the strong Galactic diffuse emission at its
  position a definitive association cannot be established. No
  statistically significant pulsations or modulation at the orbital
  period are detected. For a presumed detection, the implied
  $\gamma$-ray luminosity is $\lesssim$5\% of its spin-down
  power. This indicates that PSR J1723--2837 is either one of the
  least efficient $\gamma$-ray producing millisecond pulsars or, if
  the detection is spurious, the $\gamma$-ray emission pattern is not
  directed towards us.
\end{abstract}


\section{INTRODUCTION}
PSR J1723--2837 is a nearby ($D=750$ pc), binary, radio millisecond
pulsar (MSP) with a 1.86-ms spin period discovered by
\citet{Faulk04} in the Parkes Multibeam survey. Follow-up observations
with the Parkes, Green Bank, and Lovell telescopes have allowed
a positional localization to better than 1$''$ and detailed
parametrization of the binary orbit. Infrared, optical, and
ultraviolet spectrophotometric studies have yielded complementary
constraints on the properties of the companion star. The MSP follows
an almost circular 14.8 hour orbit, about a non-degenerate companion
star of spectra type G5 with mass $0.4-0.7$ M$_{\odot}$
\citep{Craw13}. The pulsar is rarely detected at low frequencies,
while it goes undetected at high radio frequencies for $\sim$15\% of
the orbit when the companion is generally closer to the Earth than the
pulsar, suggesting that eclipses are responsible for the
non-detections. This assertion is supported by the presence of
peculiar orbital period irregularities in the radio timing residuals,
which suggest strong tidal interactions between the neutron star and
an extended and likely mass-losing companion. Occasionally, the pulsar
goes undetected throughout the orbit (even when it is at the closest
position with respect to the observer), indicating that, at times, the
pulsar is completely enshrouded by matter released by its companion.

Its properties make PSR J1723--2837 a member of a growing class of
eclipsing MSPs termed ``redbacks'', with low-mass main-sequence-like
companions, found in both globular clusters and the field of the
Galaxy, \citep[e.g.,][]{DAm01,Arch09}.  The nature of their companions
and the irregular eclipses and rapid dispersion measure fluctuations
make them distinct from the so-called ``black widow'' eclipsing
pulsars \citep{Fru88}, which are bound to very low mass companions
($\lesssim$0.05 M$_{\odot}$).  In this sense, PSR J1723--2837 appears
similar to PSR J1023+0038, which is believed to still be transitioning
from a low-mass X-ray binary (LMXB) to a fully ``recycled'' millisecond radio
pulsar \citep{Arch09,Pat13}.  The discovery of back-and-forth switching
between accretion- and rotation-powered states of PSR J1824--2452I in
the globular cluster M28 \citep{Pap13} strongly supports the claim
that redback systems are recently activated radio MSPs, but which
may still sporadically revert to an accreting X-ray binary state. At a
distance $\sim$750 pc, based on its dispersion measure and the NE2001
model \citep{Cord02}, and confirmed by optical spectroscopy
\citep{Craw13}, PSR J1723--2837 is the nearest such system known,
nearly a factor of two closer than PSR J1023+0038 \citep[$D=1.36$
  kpc;][]{Del12}.  As such, it is a well-suited target for studies of
various aspects of these peculiar systems.

In X-rays, redback systems typically exhibit predominantly non-thermal
emission that is strongly modulated at the binary period
\citep{Bog05,Bog10,Arch10,Bog11a,Bog11b,Gen13}. This radiation is
likely produced by interaction of the energetic wind from the pulsar
with material from the companion star. The large-amplitude flux
variability likely arises due to a geometric occultation of the
X-ray-emitting region by the secondary star \citep{Bog11b}. Studing
this radiation can provide a valuable diagnostic of the physics and
geometry of MSP winds, the interaction of the two stars, and
collisionless shocks, in general.

The \textit{Fermi} Large Area Telescope (LAT) has revealed that many
MSPs are bright $\gamma$-ray sources, disproportionately accounting
for 46 of the 132 pulsars detected in pulsed
$\gamma$-rays\footnote{See
  \url{https://confluence.slac.stanford.edu/display/GLAMCOG/\\Public+List+of+LAT-Detected+Gamma-Ray+Pulsars}
  for an up-to-date list.} \citep{Abdo13}, including four of the six redbacks
currently known \citep{Hes11,Kap12,Ray12}.  As reported in
\citet{Craw13}, PSR J1723--2837 is not positionally coincident with a
catalogued $\gamma$-ray source and no pulsations are detected with
simple photon folding. Nevertheless, as recent studies have shown
\citep{Ple12,Gui12}, in principle, it is still possible to detect
pulsars in $\gamma$-rays by exploiting more sophisticated analysis
techniques such as photon probability weighting.

Herein, we present \textit{XMM-Newton} European Photon Imaging Camera
(EPIC) and \textit{Chandra X-ray Observatory} Advanced CCD Imaging
Spectrometer (ACIS) observations of PSR J1723--2837. This study
provides additional insight into the physics of this peculiar variety
of binary MSPs and establishes the X-ray characteristics of this
population.  We also investigate the $\gamma$-ray emission from this
pulsar based on the presently available \textit{Fermi} LAT data. The
work is outlined as follows. In \S 2, we detail the observations, data
reduction, and analysis procedures. In \S3, we investigate the
orbital-phase dependent variability of PSR J1723--2837 in X-rays. In
\S 4 we present phase-averaged and phase-resolved X-ray spectroscopy,
while in \S5 we describe the X-ray imaging analysis. In \S6 we attempt
to constrain the physical properties of the intrabinary shock based on
the X-ray data. In \S7 we summarize the results of a \textit{Fermi}
LAT analysis of the pulsar. We offer conclusions in \S8.

%
%
\begin{figure}[!t]
\begin{center}
\includegraphics[width=0.48\textwidth]{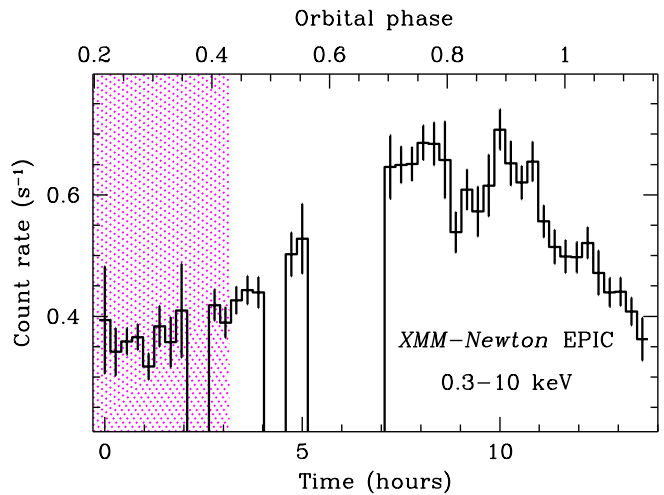}
\end{center}
\caption{Background-subtracted \textit{XMM-Newton} EPIC lightcurve of
  PSR J1723--2837 in the 0.3--10 keV band binned in 1 ks intervals.
  The gaps in the data are intervals that have been removed due to
  excessive background flaring. The dotted magenta band represents the
  approximate portion of the orbit where the pulsar undergoes a radio
  eclipse at 2 GHz.  The orbital phase is defined such that the
  companion star is between the pulsar and observer at $\phi_b=0.25$.}
\end{figure}

%
%
\begin{figure}[!t]
\begin{center}
\includegraphics[width=0.48\textwidth]{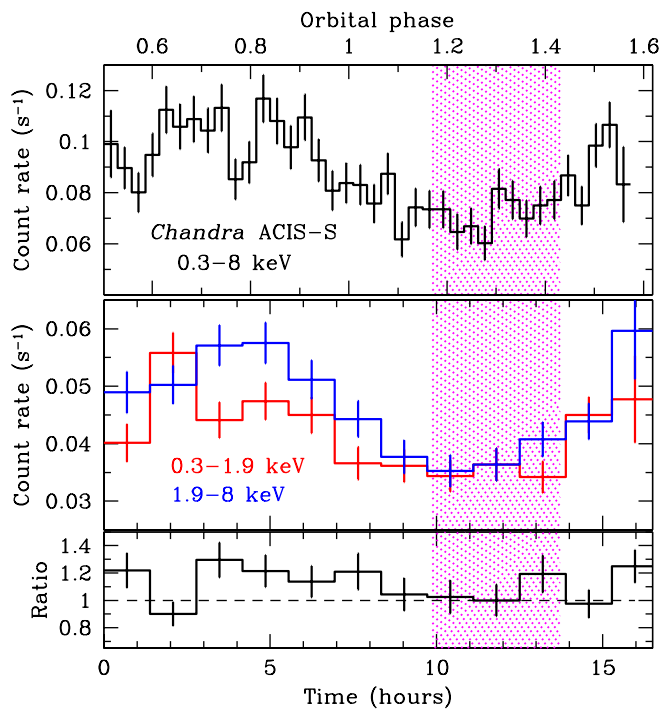}
\end{center}
\caption{\textit{Chandra} ACIS-S 0.3--8 keV (\textit{Top panel}) and
  0.3--1.9 keV and 1.9--8 keV bands (\textit{Middle panel})
  lightcurves of PSR J1723--2837 versus time and orbital phase.  The
  background contributes negligibly to the total count rate
  ($\sim$0.1\%). (\textit{Bottom panel}) The ratio obtained by
  dividing the 1.9--8 keV by the 0.3--1.9 keV lightcurve.  The radio
  eclipse interval at 2 GHz is shown by the magenta dotted band.}
\end{figure}

\section{OBSERVATIONS AND DATA REDUCTION}

\subsection{\textit{XMM-Newton}}
The \textit{XMM-Newton} observation of PSR J1723--2837 (ObsID
0653830101) was conducted on 2011 March 3 for 55 ks.  In this work, we
concentrate on the data collected with the high-throughput EPIC
instrument, which covers the 0.1--12 keV band with one pn
\citep{struder01} and two MOS \citep{turner01} CCD cameras.  All
detectors used the medium-thickness optical blocking filters and were
operated in Full Window mode.

The data were processed using version 11 of the \emph{XMM-Newton}
Science Analysis Software (SAS\footnote{The \textit{XMM-Newton} SAS is
  developed and maintained by the Science Operations Centre at the
  European Space Astronomy Centre and the Survey Science Centre at the
  University of Leicester.}) and standard screening criteria were
applied (selecting only 1- and 2-pixel events using the defaut flag
masks). The observation was affected by multiple intense soft proton
flares. The corresponding periods of high particle background were
excluded using intensity filters, following the method by
\citet{deluca04}. This resulted in a net exposure time of 23.2 ks in
the pn, 31.7 ks in the MOS\,1, and 33.2 ks in the MOS\,2.

For the timing and spectral analysis, source events were extracted
from each detector within a circular region with 40$''$ radius, which
contains $\sim$90\% of the point source energy. The background counts
were extracted from source-free regions on the same chip as the
target. The ancillary response files and the spectral redistribution
matrices were generated with the SAS scripts {\tt arfgen} and {\tt
  rmfgen}, respectively.  The spectra were binned so as to have a
minimum of 30 counts per energy channel. For the variability analysis
the data were barycentred using the DE405 ephemeris.

\subsection{\textit{Chandra}}

The \textit{Chandra} dataset was acquired on 2012 July 11 (ObsID
13713) in a continuous 49-ks effective exposure, covering 1.03 orbits
of the pulsar. The radio pulsar position was at the aim point of the
ACIS-S3 CCD set up in VFAINT mode and in a $1/8$ sub-array
configurationto ensure that the effect of photon pileup
\citep{Davis01} is minimal.  The re-processing, reduction, and
analysis of the \textit{Chandra} data were performed using
CIAO\footnote{Chandra Interactive Analysis of Observations, available
  at \url{http://cxc.harvard.edu/ciao/}} 4.4 \citep{Fruscione06} and
the corresponding calibration products (CALDB 4.4.10).  To facilitate
the identification of nebular X-ray emission, we reprocessed the level
1 data products opting for no pixel randomization and used the
background cleaning procedure appropriate for the VFAINT
mode. However, since this algorithm can reject genuine source photons
for relatively bright sources, no background cleaning was applied to
the data used for the spectroscopy and varibility studies.

For the purposes of the investigations presented in \S3 and \S4, we
extracted photons within 2$\arcsec$ of the source.  To permit spectral
fitting, the source counts in the 0.3--8 keV band were combined such
that at least 15 counts per energy bin were obtained. The background
spectrum was obtained from regions near the pulsar that are devoid of
point sources.  For the purposes of the variability study, the event
times of arrival were shifted to the solar system barycenter assuming
DE405 JPL solar system ephemeris.

The spectroscopic analyses of both the \textit{XMM-Newton} and
\textit{Chandra} observations were carried out using
XSPEC\footnote{Available at
  \url{http://heasarc.nasa.gov/docs/xanadu/xspec/index.html}.}
12.7.1. The coarse time resolution of the observations (0.4 s for
\textit{Chandra} ACIS-S, 2.6 s and 0.73 s for \textit{XMM-Newton} EPIC
MOS and pn, respectively) does not permit a search for pulsations at
the pulsar spin period.

\subsection{{\it Fermi} LAT}

For the $\gamma$-ray analysis, we retrieved Pass7
\textit{Fermi} Large Area Telecope (LAT) event data from 2008 August 4
and 2013 June 11 within 20$^{\circ}$ of the pulsar position and
accompanying spacecraft data. The analysis was carried out using the
\textit{Fermi} Science
Tools\footnote{\url{http://fermi.gsfc.nasa.gov/ssc/data/analysis/scitools/overview.html}}
v9r27p1. Following the recommended guidelines from the Fermi Science
Support Center, the data were filtered for ``source'' class events in
good time intervals with energies above 100 MeV, zenith angles smaller
than 100$^{\circ}$, and telescope rocking angles $\le$52$^{\circ}$
using the {\tt gtselect} and {\tt gtmktime} tools. The
spatial/spectral binned likelihood analysis was conducted using the
{\tt gtlike} tool based on the input counts, exposure, and source
maps, livetime cube and source model generated with the \textit{Fermi}
Science Tools.

Using the spectral parameters obtained from the likelihood analysis,
the tool {\tt gtsrcprob} was used to assign each event a probability
that it originated from PSR J1723--2837 based on the fluxes and
spectra obtained from the likelihood analysis. The {\tt gtdiffrsp} was
used to compute the integral over solid angle of a diffuse source
model convolved with the instrumental response function. Only photons
with probabilities $\ge$0.05 of being associated with the pulsar were
folded. To fold the data we used the {\tt fermi} plugin for {\tt
  tempo2} and the radio ephemeris presented in \citet{Craw13}.

%
%
\begin{figure}[!t]
\begin{center}
\includegraphics[width=0.48\textwidth]{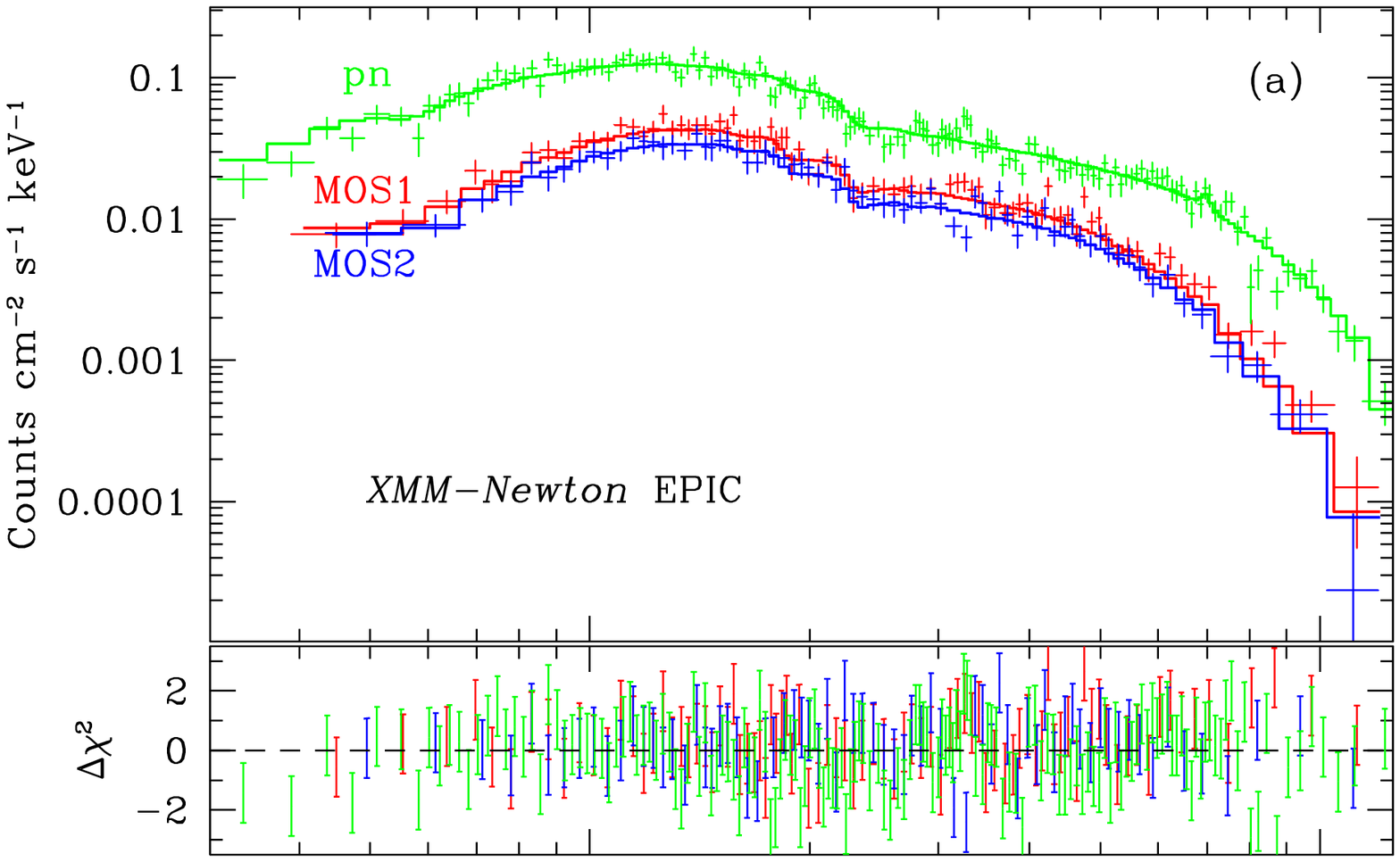}
\includegraphics[width=0.48\textwidth]{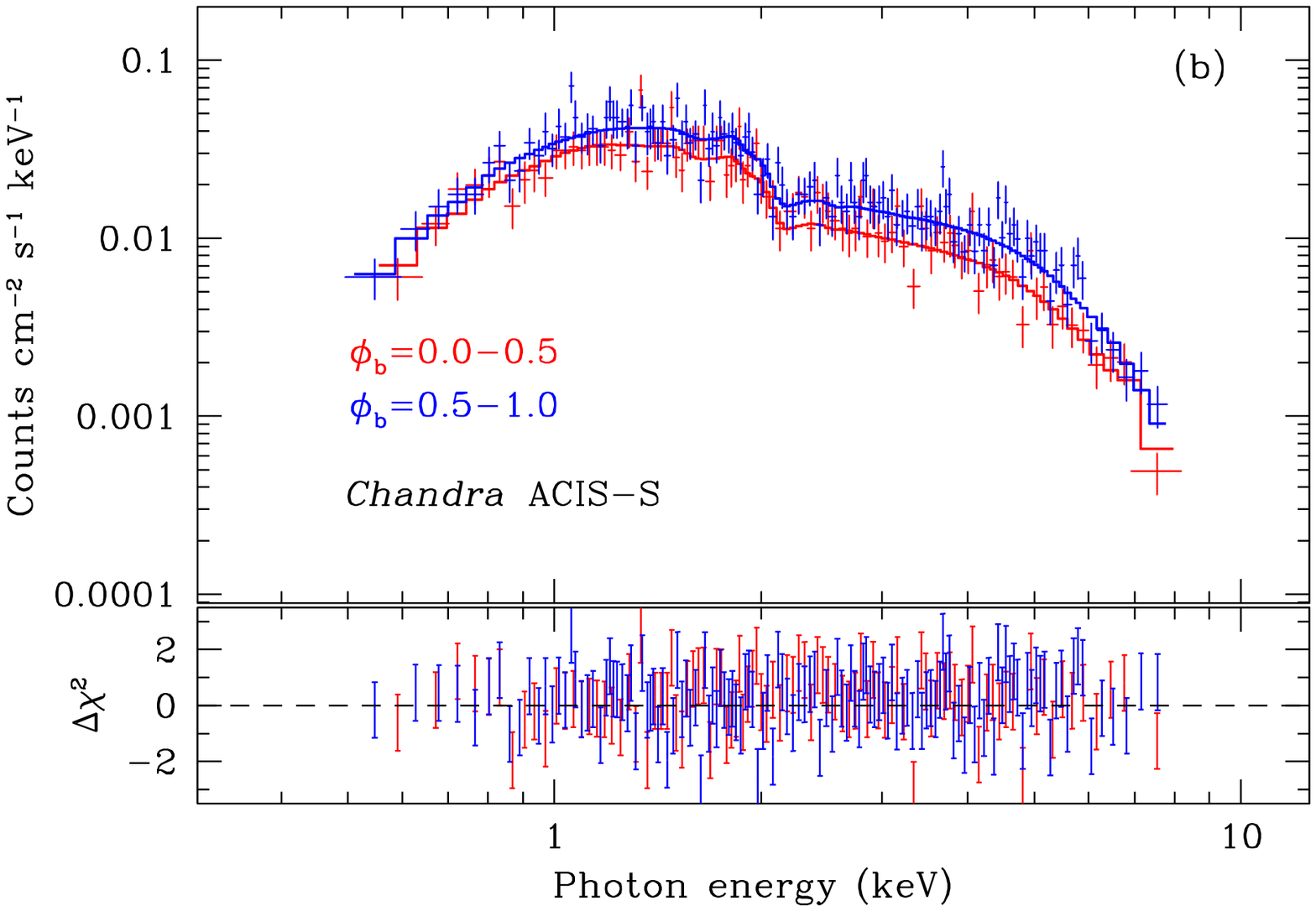}
\end{center}
\caption{(a) \textit{XMM-Newton} EPIC MOS1/2 (\textit{red} and
  \textit{blue}) and pn (\textit{green}) orbital phase-averaged X-ray
  spectra fitted with an absorbed power-law.  (b) Orbital
  phase-resolved \textit{Chandra} X-ray spectral continua of
  PSR J1723--2837 for phases $\phi_b=0.0-0.5$ (\textit{blue})
  and $\phi_b=0.5-1.0$ (\textit{red}), fitted with an absorbed
  power-law. The lower panels shows the best-fit residuals in terms of
  $\sigma$ with error bars of size one. See text and Table 1 for
  best-fit parameters.}
\end{figure}

\section{X-ray Orbital Variability}

Using the radio timing ephemeris of PSR J1723--2837 (Crawford et
al.~2013) we have determined the orbital phases of the barycentered
\textit{XMM-Newton} and \textit{Chandra} source photons. Each
observation cover approximately a single orbit. Large-amplitude flux
variability as a function of time is clearly evident in both data sets
(Figures 1 and 2).  Although substantial segments of the
\textit{XMM-Newton} data are removed due to strong flaring, the trend
in the flux modulation is still apparent. A substantial decrease (by a
factor of $\sim$2-3) in flux at superior conjunction
($\phi_b\approx0.25$) is apparent. This indicates that the X-ray flux
varies as a function of the binary period (Figures 1 and 2), a
behavior similar to what is observed analogous MSP systems, especially
PSRs J1023+0038 \citep{Arch10,Bog11b} in the field of the Galaxy and
J0024--7204W in the globular cluster 47 Tuc
\citep{Camilo00,Freire03,Bog05}.

A $\chi^2$ test on the data, binned as in Figures 1 and 2, indicates
negligible probabilities of $4\times10^{-94}$ (20.5$\sigma$) and
$2\times10^{-14}$ (7.6$\sigma$) that the observed flux variability
arises from a constant flux distribution for \textit{XMM-Newton} and
\textit{Chandra}, respectively.  A more robust estimate is obtained
from the Kuiper test \citep{Pal04}, which considers the unbinned
lightcurves, weighted to account for the non-uniform exposure across
the orbit. This approach gives probabilities of $6\times10^{-101}$
(21.3$\sigma$) and $2.8\times10^{-21}$ (9.4$\sigma$), for the
\textit{XMM-Newton} and \textit{Chandra} data respectively, that
events being drawn from a constant distribution would exhibit this
level of non-uniformity.  There is no appreciable spectral variability
as a function of orbital phase (bottom panel of Figure 2). There is
only marginally significant evidence for spectral hardening between
phases $\sim$0.7--0.9, although as discussed below, pile-up in the
\textit{Chandra} data may be partly responsible for this.

\begin{deluxetable*}{lcccccccc}
\tablecolumns{9} 
\tablewidth{0pc}
\tablecaption{Summary of X-ray Spectroscopy for PSR J1723--2839.}
\tablehead{ \colhead{}  & \colhead{\textbf{\textit{XMM}}} & \colhead{} & \colhead{\textbf{\textit{XMM+CXO}}} &  \colhead{} & \multicolumn{4}{c}{\textbf{\textit{CXO}}} \\
\cline{2-2} \cline{4-4} \cline{6-9} \\ 
\colhead{} & \colhead{Total} & \colhead{} & \colhead{Total} & \colhead{} & \colhead{Total} & \colhead{$\phi_{b,1}$} & \colhead{$\phi_{b,2}$} & \colhead{Joint} \\
Model\tablenotemark{a} & \colhead{ } &  \colhead{ } &  \colhead{ } & \colhead{ }  & \colhead{ } & \colhead{$(0.0-0.5)$} & \colhead{$(0.5-1.0)$} & \colhead{$\phi_{b,1}+\phi_{b,2}$}}
 \startdata
\textbf{Power-law} & 	&	& &	&  &  & &	\\
\hline
$N_{\rm H}$ ($10^{21}$ cm$^{-2}$) 	&  $2.0^{+0.09}_{-0.09}$ &   & $1.9^{+0.08}_{-0.08}$ & & $1.6^{+0.2}_{-0.2}$	& $1.8^{+0.4}_{-0.3}$	& $1.3^{+0.3}_{-0.2}$	&  $1.5^{+0.2}_{-0.2}$	\\
$\Gamma$	 &  $1.15^{+0.02}_{-0.02}$  & & $1.13^{+0.02}_{-0.02}$  & & $1.00^{+0.04}_{-0.04}$	& $1.10^{+0.07}_{-0.07}$	& $0.88^{+0.06}_{-0.06}$	& $1.13^{+0.05}_{-0.05}/0.85^{+0.05}_{-0.04}$ \\
$F_X$ (0.3--8 keV)\tablenotemark{c} & $1.87^{+0.02}_{-0.02}$ &  & $1.87^{+0.02}_{-0.02}/1.27^{+0.02}_{-0.02}$  & & $1.29^{+0.02}_{-0.02}$  	& $1.06^{+0.03}_{-0.03}$	& $1.48^{+0.04}_{-0.04}$	& $1.05^{+0.03}_{-0.03}/1.49^{+0.03}_{-0.04}$	\\
$\chi^2_{\nu}$/dof    &  $1.04/321$  &   & $1.02/519$ & & $0.97/199$ & $1.00/94$	& $0.95/129$	&  $0.98/225$	\\
\hline
\textbf{Power-law + NSA}\tablenotemark{b}	& & 	&	&	&  & &	& \\
\hline
$N_{\rm H}$ ($10^{21}$ cm$^{-2}$)	 &  $3.4^{+0.7}_{-0.8}$   &  & $2.1^{+0.3}_{-0.2}$ &  & $0.96^{+0.42}_{-0.93}$  & $5.0^{+1.1}_{-1.6}$ & $4.0^{+1.3}_{-2.2}$	& $0.80^{+0.58}_{-0.72}$	\\
$\Gamma$                        &  $1.15^{+0.04}_{-0.04}$  &  & $1.12^{+0.02}_{-0.02}$ & & $0.77^{+0.26}_{-0.68}$ & $1.33^{+0.10}_{-0.12}$	& $0.97^{+0.10}_{-0.11}$	& $0.82^{+0.22}_{-0.54}$ 	\\
$T_{\rm eff}$ ($10^6$ K)  &   $0.82^{+0.30}_{-0.17}$         &  & $1.10^{+0.75}_{-0.27}$ &  & $<$$6.59$	& $0.32^{+0.27}_{-0.41}$ & $0.60^{+0.74}_{-0.13}$ 	& $0.50^{+0.22}_{-0.08}$ 	\\
$R_{\rm eff}$ (km)  & $4.3^{+9.9}_{-4.1}$  &  &  $<$$0.32$ & & $<$$0.05$    & $115^{+201}_{-112}$ 	& $11.6^{+31.3}_{-11.5}$ 	&  $14.5^{+30.3}_{-14.2}$	\\
Non-Thermal fraction\tablenotemark{d} & $0.99^{+0.01}_{-0.04}$   &    &	$0.98^{+0.02}_{-0.12}/0.97^{+0.3}_{-0.12}$ & & $0.93^{+0.03}_{-0.03}$ & $0.41^{+0.06}_{-0.08}$ 	& $0.84^{+0.05}_{-0.04}$ 	& $0.84^{+0.08}_{-0.07}/0.89^{+0.07}_{-0.06}$ 	\\
$F_X$ (0.3--8 keV)\tablenotemark{c} & $2.23^{+0.04}_{-0.03}$ & & $1.91^{+0.04}_{-0.03}/1.29^{+0.03}_{-0.03}$ & & $1.41^{+0.05}_{-0.07}$ 	& $2.90^{+0.04}_{-0.04}$ & $1.82^{+0.05}_{-0.07}$ & $1.06^{+0.06}_{-0.07}/1.49^{+0.06}_{-0.06}$	\\
$\chi^2_{\nu}/$dof    & $1.01/316$   &  & $1.03/517$ & & $0.98/197$ & $0.98/92$ & $0.95/127$	& $0.98/222$ \\
\hline
\textbf{Power-law + MEKAL}\tablenotemark{e} &	& 	&  &	&	&   &	& \\
\hline
$N_{\rm H}$ ($10^{21}$ cm$^{-2}$)	& $2.1^{+0.2}_{-0.1}$  &    & $2.0^{+0.2}_{-0.1}$ &  & $2.2^{+0.6}_{-0.5}$  & $3.6^{+1.4}_{-1.2}$ & $1.3^{+0.3}_{-0.3}$	& $0.19^{+0.6}_{-0.5}$	\\
$\Gamma$        &  $1.14^{+0.03}_{-0.02}$  & 	  & $1.13^{+0.02}_{-0.02}$  &  & $1.05^{+0.06}_{-0.06}$	& $1.23^{+0.11}_{-0.11}$	& $0.85^{+0.07}_{-0.07}$	& $1.08^{+0.07}_{-0.07}/0.95^{+0.07}_{-0.07}$ 	\\
$kT$ (keV)      & $0.57^{+0.14}_{-0.27}$ &        & $0.32^{+0.12}_{-0.06}$  & & $0.25^{+0.13}_{-0.05}$	& $0.24^{+0.06}_{-0.03}$ & $<1.41$ 	& $0.29^{+0.19}_{-0.08}$ 	\\
Non-thermal fraction\tablenotemark{d}   & $0.99^{+0.01}_{-0.04}$  &   &	$0.99^{+0.01}_{-0.03}/0.98^{+0.02}_{-0.03}$ & & $0.97^{+0.03}_{-0.03}$ & $0.88^{+0.06}_{-0.08}$ 	& $0.99^{+0.01}_{-0.06}$ 	& $0.98^{+0.02}_{-0.12}/0.98^{+0.02}_{-0.12}$ 	\\
$F_X$ (0.3--8 keV)\tablenotemark{c} & $1.89^{+0.02}_{-0.02}$  &	& $1.91^{+0.04}_{-0.03}/1.29^{+0.03}_{-0.03}$ &  & $1.34^{+0.11}_{-0.02}$ 	& $1.29^{+0.06}_{-0.05}$ & $1.49^{+0.04}_{-0.04}$ & $1.09^{+0.07}_{-0.07}/1.53^{+0.06}_{-0.06}$	\\
$\chi^2_{\nu}/$dof     & $1.03/316$ &    & $1.02/517$ & & $0.97/197$ & $0.99/92$ & $0.95/127$	& $0.98/222$
\enddata 
\tablenotetext{a}{All quoted uncertainties and limits are at a 1$\sigma$ confidence level.}
\tablenotetext{b}{For the {\tt nsa} model, a star with $R=10$ km, $M=1.4$ M$_{\odot}$ is assumed. The un-redshifted effective radius, $R_{\rm eff}$, was computed assuming $D=750$ pc.}
\tablenotetext{c}{Unabsorbed X-ray flux (0.3--8 keV) in units of
  $10^{-12}$ ergs cm$^{-2}$ s$^{-1}$.}
\tablenotetext{d}{Fraction of unabsorbed flux from the nonthermal component in the 0.3--8 keV band.} 
\tablenotetext{e}{For the MEKAL model, solar abundances are assumed.}
\end{deluxetable*}

\section{Phase-averaged X-ray Spectroscopy}

In similar MSP binary systems, the phase-integrated X-ray continuum is
well represented by a power-law, while a single thermal (either
blackbody or neutron star hydrogen atmosphere) model fails to
reproduce the spectral shape. In some instances, an acceptable fit is
obtained with a composite power-law plus thermal model.  Like many
MSPs studied in X-rays \citep{Zavlin06,Bog06,Bog09}, PSR J1723--2837
is expected to have hot polar caps due to a return current of
particles from the pulsar magnetosphere. Based on this, in addition to
a single-component power-law we consider a composite model consisting
of an absorbed power-law and a NS atmosphere.  We choose the NSA
atmosphere model \citep{Zavlin96} instead of a blackbody since it has
been shown that the thermal pulsations from the nearest MSPs favor an
atmosphere \citep{Zavlin98,Bog07,Bog09,Bog13}, as expected at the
surface of objects ``recycled'' via accretion of matter. A fraction of
the X-rays associated with J1723--2837 could arise from a thermal
plasma within or around the binary, possibly from the active corona of
the companion star or intra-binary plasma that causes the radio
eclipses. Hence, we also present fits using a a power-law plus MEKAL
hot diffuse plasma model, which includes line emissions from several
elements based on input metal abundances \citep{Mewe85,Mewe86,Lie95}.
Table 1 summarizes the results using the three different models.

A joint fit to the \textit{XMM-Newton} and \textit{Chandra}
phase-averaged spectra results in statistically unacceptable fits due
to a significant flux difference between the two data sets. This
discrepancy cannot be explained by the partial orbital coverage of the
\textit{XMM-Newton} data.  Given the $1.3$ year separation of the two
observations, it is likely the result of long term flux variations
from the X-ray-emitting intrabinary shock, possibly due to changes in
the stellar activity of the secondary star or fluctuations in the
outflow of gas from the companion through the inner Lagrangian
point. Similar flux changes on timescales of years have also been
observed in the ``redback'' PSR J0024--7204W in the globular cluster
47 Tuc \citep{Bog05,Cam07}, suggesting this may be a common feature of
these systems.  As a consequence of the appreciable flux difference,
we have fitted the two observations jointly but with independent flux
normalizations as well as separately.  From a statistical standpoint,
the simple pure power-law and power-law plus thermal component both
produce satisfactory fits.  The addition of a NSA or MEKAL component
results in a slight improvement in the quality of the fit compared to
a pure power-law but is not statistically significant.  The parameters
of the NSA and MEKAL components are poorly constrained as their
contribution to the total flux is typically only a few percent.  The
hydrogen column density through the Galaxy along the line of sight to
the pulsar is $\sim$$3.8\times10^{21}$ cm$^{-2}$ based on
\citet{Kalb05}. The best fit values of $N_{\rm H}$ \citep[assuming
  abundances from][]{And89} in Table 1 are generally consistent with
this value.

\subsection{Orbital Phase-resolved Spectrum}
In the case of the \textit{XMM-Newton} data, the large gaps in orbital
phase coverage resulting from background flare removal are not
conducive to phase-resolved spectroscopy.  Therefore, we only consider
the \textit{Chandra} observation for this analysis.  We divide the
data over two orbital phase intervals: $\phi_b=0.0-0.5$ (around the
minimum in X-ray flux) and $\phi_b=0.5-1.0$. The two spectra were
fitted both separately and jointly. For the latter, for the thermal
component we tied both the temperature and effective radius in all
instances since the emission associated with the pulsar is not likely
to exhibit any variability as a function of orbital phase.  The
best-fit parameters of the orbital phase-resolved spectral fits are
listed in of Table 1.

A simple power-law provides a very good fit to the total
phase-averaged X-ray spectrum, as well as each of the spectra at
phases $0-0.5$ and $0.5-1$. The same is true for the joint fit
of phase-resolved spectra.  In all instances, the power-law around
flux maximum is significantly harder compared to flux minimum and
inferred from the phase-averaged \textit{XMM-Newton} data.  Although
this could be indicative of intrinsic spectral hardening of the
source, it could also arise due to photon pile-up on the ACIS-S
detector.  Pile-up occurs when two or more photons, arriving at the
detector during one frame time, are erroneously identifed as a single
photon with the sum of the photon energies or else discarded (Davis
2001). The result is a distortion of the intrinsic shape of the source
spectrum, causing an artificial hardening of the spectrum.  Using
PIMMS\footnote{Available at
  \url{http://cxc.harvard.edu/toolkit/pimms.jsp}.}, we find that the
predicted pile-up for this source in the $1/8$ sub-array mode of
ACIS-S is $\sim$2\% based on the phase-averaged count rate.  Indeed,
fitting a power-law spectrum with the {\tt pileup} model in {\tt
  XSPEC} yields a slightly steeper power-law with
$\Gamma=1.05^{+0.05}_{-0.07}$ for the phase-avaraged ACIS-S spectrum
and $\Gamma=0.94^{+0.07}_{-0.05}$ for the orbital phase interval
$\phi_b=0.5-1.0$, which is more consistent with the results from the
other fits.

\section{X-ray Imaging Analysis}

PSR J1723--2837 is by far the brightest source in the
\textit{XMM-Newton} EPIC and \textit{Chandra} ACIS-S images (Figure
4). This confirms that the \textit{ROSAT} X-ray point source 1RXS
J172323.7--283805, located 13$''$ from the position of the pulsar, is
in fact PSR J1723--2837, as suggested in \citet{Craw13}.  The X-ray
position measured from the \textit{Chandra} image using {\tt
  wavdetect} is $\alpha_X=17^{\rm h}23^{\rm m}23\fs19$,
$\delta_X=-28^{\circ}37\arcmin57\farcs49$, which differs from the
radio position $\alpha_r=17^{\rm h}23^{\rm m}23\fs1856$,
$\delta_r=+17^{\circ}37\arcmin57\farcs17$ \citep{Craw13} by just
$+0.06\arcsec$ and $-0.32\arcsec$ in right ascension and declination,
respectively, significantly smaller than the uncertainty in the
absolute astrometry of \textit{Chandra}\footnote{See
  \url{http://cxc.harvard.edu/cal/ASPECT/celmon/}}.

The sub-arcsecond angular resolution afforded by \textit{Chandra}
allows us to look for nebuar X-ray radiation surrounding the pulsar.
Redback systems are of particular interest in this regard since any
information about recent accretion could, in principle, be encoded in
any anomalies in the diffuse X-rays associated with the pulsar.
However, for $\gtrsim$2$\arcsec$ away from PSR J1723--2837, there is
no indication for excess emission in any direction away from the
pulsar due to a bow shock or wind nebula. To formally verify this, we
generated 20 simulated observations of PSR J1723--2837 with
ChaRT\footnote{The Chandra Ray Tracer, available at
  \url{http://cxc.harvard.edu/soft/ChaRT/cgi-bin/www-saosac.cgi}} and
MARX 4.5\footnote{Available at
  \url{http://space.mit.edu/cxc/marx/index.html}.}, using the
specifics of the \textit{Chandra} observation and the best-fit X-ray
spectrum as input. The average of the simulated point spread functions
was subtracted from the observed image to identify any residual
emission relative to the background level.  For $\lesssim$1$\arcsec$
from the center, the difference image reveals appreciable residuals
(both positive and negative) that are azimuthally asymmetric .  In
principle, a shift between the position reported by {\tt wavdetect}
and the true source position could produce such residuals. To
investigate this possibility, we have repreated the analysis using a
range of shifts along the azimuthal direction in which the residuals
are most significant in an attempt to minimize them. None yielded an
improvement in the difference image.  Based on this, the most likely
explantion is that these residuals arise due to imperfections in the
model of the High-Resolution Mirror Assembly optics
\citep{Juda10}\footnote{See also
  \url{http://hea-www.harvard.edu/\~juda/memos/\\HEAD2010/HEAD2010\_poster.html}
  and
  \url{http://cxc.harvard.edu/cal/Hrc/PSF/acis\_psf\_2010oct.html}}.
Beyond $\sim$1$\arcsec$ of the pulsar the difference image does not
show any deviations from what is expected from a point source (Figure
5), confirming the absence of any diffuse emission surrounding the
pulsar.

Based on the background count rate around the pulsar ($2\times10^{-6}$
counts s$^{-1}$ arcsec$^{-2}$ for $0.3-6$ keV) and taking a typical
power-law spectrum for X-ray PWNe \citep{Kar08} with $\Gamma=1.5$, the
limit on the pulsar wind nebula luminosity is
$\sim$$1\times10^{29}$ ergs s$^{-1}$ for a distance of $750$ pc.

%
\begin{figure}[t]
\begin{center}
\includegraphics[width=0.48\textwidth]{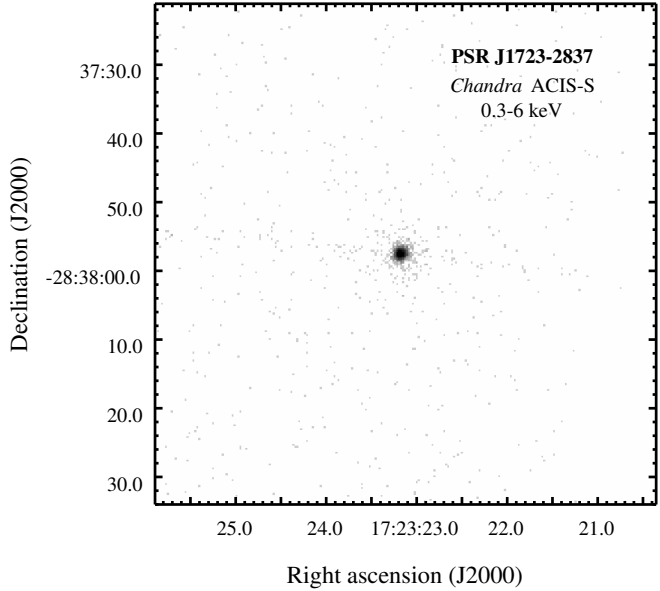}
\end{center}
\caption{\textit{Chandra} ACIS-S3 J2000 $1.2\arcmin\times1.2\arcmin$
  image binned at the intrinsic resolution of the ACIS-S detector
  ($0.5\arcsec$) showing the X-ray counterpart to PSR J1723--2837 in
  the 0.3--6 keV interval. The grey scale represents counts increasing
  logarithmically from 0 (white) to 488 (black).}
\end{figure}

%
%
\begin{figure}[t]
\begin{center}
\includegraphics[width=0.48\textwidth]{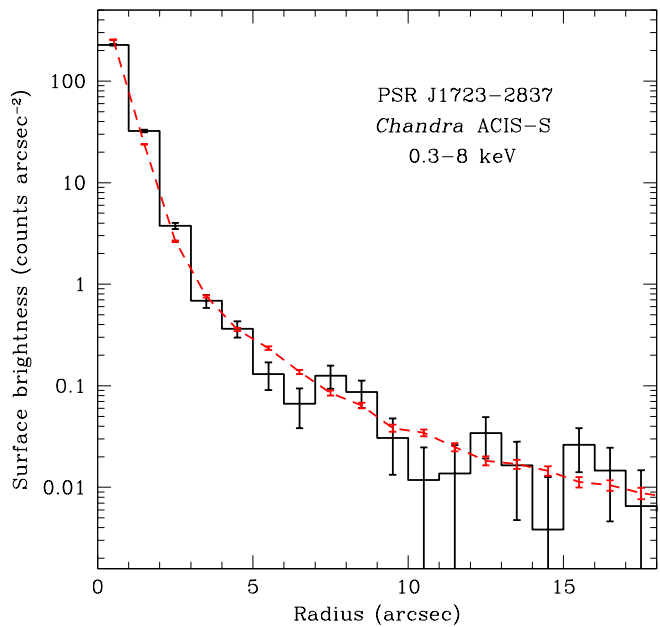}
\end{center}
\caption{Radial profile of the observed ACIS-S point spread function
  (PSF) of the X-ray counterpart of PSR J1723--2837 after background
  subtraction (\textit{histogram}) and the mean of 20 synthetic PSFs
  (\textit{red dashed line}).}
\end{figure}

\section{Probing the Physics of the Intrabinary Shock}
In \citet{Bog11b}, simple geometric modeling of PSR J1023+0038
revealed that an intrabinary shock localized primarily near the inner
Lagrangian point and/or at the face of the companion can naturally
account for the X-ray variability. In this interpretation, the
obstruction of the observer's view of the X-ray region by the
secondary causes the apparent decline in flux at $\phi\approx
0.25$. Given the qualitative similarities in observed X-ray
properties, it is highly probable that the location of the shock in
J1723--2837 is the same.

Using this information it is possible to gain quantitative insight
into the properties of the intrabinary shock. Due the compactness of
the binary ($a\approx 2\times10^{11}$ cm), it is likely that the shock
is formed in a relatively intense magnetic field, meaning that
synchrotron produced by accelerated particles is the most probable
X-ray emission mechanism.  The resulting synchrotron luminosity is a
function of the strength of the post-shock field and the ratio between
the Poynting flux and particle flux $\sigma$ (the wind magnetization
factor). If the wind is dominated by kinetic energy, $\sigma \approx
0.003$ \citep[like in the Crab nebula][]{Ken84}, while for a
magnetically dominated wind, $\sigma \gg 1$. Based on the prescription
presented by \citet{Arons93}, the field strength immediately past the
shock is defined by
$B_1=[\sigma/(1+\sigma)]^{1/2}(\dot{E}/cf_{p}r^2)^{1/2}$, where $f_p$
defines the portion of the sky into which the pulsar wind is emitted,
while $r$ is the separation between the pulsar and the shock
front. For PSR J1723--2837, the approximate distance from the MSP to
L$_1$ assuming $M_{\rm MSP}=1.4$ M$_{\odot}$ and $i=37^{\circ}$ is
$r\approx2\times10^{11}$ cm.  For a pulsar wind that is emitted
isotropically ($f_{p}=1$) with $\dot{E}=4.6\times10^{34}$ ergs
s$^{-1}$ this produces $B_1\approx0.68$ G ($\sigma=0.003$) and
$B_1\approx 12$ G ($\sigma\gg1$).  The resulting magnetic field
strength past the shock is $B_2=3B_1\sim 2$ G or $B_2\sim37$ G,
respectively.  In order to produce photons with energies
$\varepsilon_{\rm keV}=0.3-8\sim1$ keV by synchrotron radiation,
relativistic $e^{pm}$ with Lorentz factors
$\gamma=2.4\times10^5(\varepsilon/B_2)^{1/2}$ (with $\varepsilon$ in
units of keV and $B_2$ in G) are required, which yields
$\sim$$0.2\times 10^5$ ($\sigma\gg1$) and $\sim$$1\times10^5$
($\sigma=0.003$).  The associated radiative loss time is then $t_{\rm
  synch}=5.1\times10^8(\gamma B_2^2)^{-1}\sim 2-145$ s \citep{Ryb79}.
Assuming a shock region that is $\sim$1 R$_{\odot}$, the radius of the
companion's Roche lobe \citep[][]{Craw13}, the residence times of the
synchrotron-emitting e$^{\mp}$ in the shock are $t_{\rm
  flow}=c/3R\approx13$ s (for $\sigma=0.003$) and $t_{\rm
  flow}=c/R\approx 40$ s (for $\sigma\gg 1$).

The luminosity from the shock due to synchrotron radiation can be
computed approximately using the expression $f_{\rm shock}
f_{\varepsilon} L_{\varepsilon}=f_{\rm synch}f_{\gamma}f_{\rm
  geom}\dot{E}$.  Here $f_{\rm synch}$ represents the radiative
efficiency of the synchrotron mechanism, $f_{\gamma}$ corresponds to
the portion of the wind power that goes into accelerating e$^{\mp}$
with Lorentz factor $\gamma$, $f_{\rm geom}$ is the portion of the MSP
outflow that encounters material from the secondary star, $f_{\rm
  shock}$ is the fraction of the total system luminosity that arises
due the shock, and $L_{\varepsilon}$ is the X-ray luminosity in the
energy interval under consideration, while $f\varepsilon$ is the
fraction of the total synchrotron spectrum falling in the observed
energy band.  Using $f_{\rm synch}=(1+t_{\rm synch}/t_{\rm
  flow})^{-1}$, we find values of $f_{\rm synch}\approx0.02$ and
$f_{\rm synch}\approx0.8$, corresponding to $\sigma=0.003$ and
$\sigma\gg 1$, respectively.  For a radius of $\sim$1 R$_{\odot}$, the
secondary encounters $f_{\rm geom} \approx 0.01$ of the pulsar's
outflow if the MSP wind is uniformly emitted in all
direction. However, it is highly probable that the MSP wind is
significantly anisotropic, with most of it flowing out equatorially
since such a geometry is observed in the Crab pulsar
\citep{Hest95,Mich94}. Given that during the LMXB accretion phase the
pulsar spin axis vector has become parallel  with the orbital
angular momentum vector, the wind
should be emitted predominantly in the orbital plane \citep{Bhatt91}.
For a wind emitted only in the orbital
plane, the companion star would intercept $f_{\rm geom}\approx0.08$ of
the total wind energy. Thus, we set $0.01<f_{\rm geom}<0.08$.

Based on the phase-resolved spectroscopic analysis, the intrinsic
shock luminosity (in the absence of eclipses) can be estimated to be
$L_{\varepsilon}\approx1\times10^{32}$ ergs s$^{-1}$ for particles
with $\gamma\approx10^5$. With $f_{\rm shock}\approx1$ as obtained from
the spectroscopic analysis, we find $27\lesssim f_{\gamma}\lesssim218$
($\sigma=0.003$) and $0.14\lesssim f_{\gamma}\lesssim1.1$
($\sigma\gg1$).  In the case of $\sigma=0.003$, the implied range of $f_{\gamma}$
is clearly unphysical as it exceeds unity, even if the shock region
receives 100\% of the pulsar wind power. This problem is further
exacerbated if the cut off of the energy spectrum is significantly
above 10 keV, such that the observed flux isonly a fraction of the
emitted flux. This would increase the fraction of the wind power that
has to go in electron acceleration.  This implies that for PSR
J1723--2837 the wind in the vicinity of the shock is probably magnetically
dominated.

In the case of $\sigma\gg1$, obtaining comparable values to that of the Crab
pulsar ($f_{\gamma}=0.04$) can be obtained by assuming that the bulk
of the X-ray-emitting region is confined to an equatorial strip (as
illustrated in Figure 6). It is interesting to note that a similar
feature is necessary to account for the peculiar He I lines seen in
the analogous MSP binary PSR J1740--5340 in  NGC
6397 \citep{Ferr03}.  This requires either a sheet-like pulsar wind
and/or a strong outflow from the companion that is preferentially
emitted along the stellar equator.  Alternatively, a realistic value
of $f_{\gamma}$ can be achieved for both scenarios if the secondary
star has a relatively high surface magnetic field, $\sim$$10^{2-3}$ G
\citep[see][and references therein]{Donati09}, at the high end of
fields measured for main sequence stars. However, for values
$\lesssim$$10^2$ G, the emission region still needs to be a factor of
$\lesssim$$10$ smaller than the secondary star.

%
%
\begin{figure}[t]
\begin{center}
\includegraphics[width=0.48\textwidth]{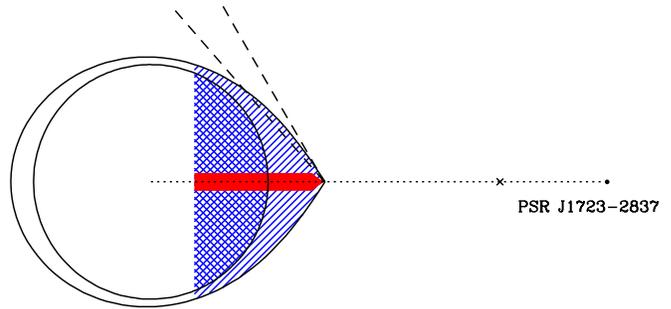}
\end{center}
\caption{Schematic illustration of the PSR J1723--2837 system. The
  blue hatched and cross-hatched regions show the portion of the
  companion that is illuminated by the pulsar wind for the case of a
  Roche-lobe filling and 90\% filling star, respectively. The red
  strip depicts an equatorial band that is 10\% of the stellar
  radius. The dotted line shows the semi-major axis and the cross
  marks the center of mass of the binary. The dashed lines delineate
  the range of possible lines of sight to the observer for
  $i=30-41^{\circ}$ at superior conjunction.}
\end{figure}

In \citet{Craw13}, it was suggested that the measured $\dot{P}$ and
hence all parameters derived from it are affected and possibly
dominated by the kinematic (Shklovskii) effect. This implies that the
pulsar's true spin-down luminosity is significantly smaller than the
derived value of $\dot{E}=4.6\times10^{34}$ erg s$^{-1}$. If we
consider a smaller intrinsic $\dot{E}$ in the calculations above, the
required X-ray conversion efficiency becomes unrealistically large,
unless we invoke a very compact X-ray-emitting shock region and/or a
companion star with a $\sim$$10^3$ G surface magnetic field.
Thus, in order to further constrain the physics and geometry of the
intrabinary shock it is important to determine the intrinsic $\dot{P}$
of the pulsar by way of continuing radio timing observations or
optical proper motion measurements.

\section{\textit{FERMI} LAT ANALYSIS}

\subsection{Binned Likelihood Analysis}

PSR J1723--2837 does not fall within the 95\% confidence region of any
source in the Fermi LAT 2-Year Point Source Catalog
\citep{Nolan12}. The two nearest published sources, 2FGL J1717.3--2809
and 2FGL J1728.0--2737c, both lie 1.4$^{\circ}$ away.  A visual
inspection of the nearly five years of \textit{Fermi} LAT data we have
retrieved reveals no obvious $\gamma$-ray source at the pulsar
position.  This is in large part due the pulsar being situated only
$4^{\circ}$ above the Galactic plane, where the diffuse Galactic
emission component is quite strong and source confusion can be
particularly problematic.

To formally establish whether PSR J1723--2837 is a $\gamma$-ray
source, we carried out a binned likelihood analysis by considering a
source at the position of the pulsar modeled by a power-law with an
exponential cutoff, with the form $dN/dt\propto E^{-\Gamma}
\exp(-E/E_c)$, where $\Gamma$ is the spectral photon index and $E_c$ is
the spectrum cutoff energy.  The parameters of the putative
$\gamma$-ray pulsar, the 44 sources within $10^{\circ}$ of the pulsar,
and the normalization factors of diffuse components were left free in
the fit. We also consider emission from sources up to $15^{\circ}$
away but keep their parameters fixed.  The source model also included
contributions from the extragalactic diffuse emission and the residual
instrumental background, jointly modeled using the {\tt
  iso\_p7v6source} template, and from the Galactic diffuse emission,
modeled with the {\tt gal\_2yearp7v6\_v0} map cube.

The likelihood analysis yields a source test statistic \citep[$TS$,
  see][ for a definition]{Nolan12} value of $57$, corresponding to a
$\sim$$7.6\sigma$ significance, for the putative pulsar $\gamma$-ray
counterpart. However, the best-fit pulsar spectrum tends towards very
steep photon indices ($\Gamma\sim3$) and low values of the cutoff
energy ($E_c\lesssim500$ MeV), likely owing to the paucity of photons
above $\sim$500 MeV.  These values are substantially different from
those of the current sample of $\gamma$-ray detected MSPs \citep[see
  Table 10 in][]{Abdo13}, which have average photon index and cutoff
energy of $\Gamma=1.3$ and $E_c=2.2$ GeV, with the lowest values being
$\Gamma=1.9$ and $E_c=1.1$ GeV.  Fixing the pulsar parameters to the
latter set of values results in $TS=45$ ($\sim$6.7$\sigma$).  To
mitigate the effect of the overwhelming Galactic diffuse background
at $\sim$100 MeV, we also carried out the likelihood analysis for
photon energies $>$300 MeV. The result is a substantially diminished
significance of the source (with $TS=22$, corresponding to
$\sim$4.7$\sigma$) and a best fit with a very steep power-law
$\Gamma\approx3$ and an abnormally low cutoff $E_c\approx200-500$
MeV. Using $\Gamma=1.9$ and $E_c=1.1$ GeV as fixed parameters results
in $TS=20$.

As the ``c'' suffix designates, the source 2FGL J1728.0--2737c is
confused, indicating that its reported source position and spectrum
are unreliable.  Based on this, we consider the possibility that this
$\gamma$-ray source is perhaps the pulsar and remove it from the input
model. The result is $TS=74.4$ ($\sim$8.6$\sigma$) for the pulsar for
$>$100 MeV but again with a very steep spectrum and a low cutoff
energy for the best fit.  In this case, fixing the parameters to
$\Gamma=1.9$ and $E_c=1.1$ GeV, results in $TS=65.3$, corresponding to
$\sim$8$\sigma$.  However, the same fit for $>$300 MeV results only in
$TS=12.7$.

Although from a statistical standpoint the pulsar is consistent with
being a $\gamma$-ray source, the veracity of the $\gamma$-ray
detection appears to depend strongly on the accuracy of the Galactic
diffuse model. The tendency towards abnormal values of $\Gamma$ and
$E_c$, as well as the dramatic decline in the source significance for
$>$300 MeV, is an indication that the bulk of $\gamma$-ray emission at
the pulsar position may not actually be associated with PSR
J1723--2837 and is possibly due to excess diffuse emission that is not
properly accounted for in the current model.  If the $\gamma$-ray
emission is in fact associated with the pulsar, the implied energy
flux from PSR J1723--2837 assuming $\Gamma=1.9$ and $E_c=1.1$ GeV is
$F_{\gamma}\approx2\times10^{-11}$ erg cm$^{-2}$ s$^{-1}$ for energies
above 100 MeV. For $D=750$ pc, this corresponds to a $\gamma$-ray
luminosity of $L_{\gamma}\approx 2\times 10^{33}$ erg s$^{-1}$. Given
the ambiguity regarding the pulsar detection, this provides a
conservative upper limit on the $\gamma$-ray production efficiency of
$\sim$5\% of the pulsar spin-down luminosity, towards the low end of
values in the current sample of \textit{Fermi} LAT MSPs \citep[see
  Table 10 in][]{Abdo13}.  This could mean that PSR J1723--2837 is
either one of the least efficient $\gamma$-ray emitting MSPs or, if
the detection is spurious, the $\gamma$-ray emission pattern is not
directed towards us \citep{Rom11}.

\subsection{Photon-Weighted Pulsation Search}

As reported in \citep{Craw13}, the timing irregularities associated
with this PSR J1723--2837 requires the addition of multiple orbital
period derivatives in order to obtain a satisfactory radio timing
solution. As a result, the best radio timing ephemeris cannot be
reliably extrapolated to fold \textit{Fermi} LAT photons over the
entire $\sim$5 year span of the mission. As a result, we restrict our
analysis to events detected in the interval over which the
pulsar ephemeris is valid (MJDs 55101.8--55803.8).

Folding the \textit{Fermi} LAT photons extracted with various energy
cuts and acceptance cone radii does not yield statistically
significant $\gamma$-ray pulsations.  However, as demonstrated in
recent studies \citep{Kerr11,Gui12} weighting the $\gamma$-ray photons
by the probability that they originate from a pulsar significantly
enhances the sensitivity to faint pulsations. This is especially
critical for PSR J1723--2837 owing to its proximity to the Galactic
plane ($b=4.2^{\circ}$), where the diffuse $\gamma$-ray background is
strong.  The resulting best fit source model from the binned
likelihood analysis was used in conjunction with the {\tt gtsrcprob}
script to assign a probability to each photon of being associated with
PSR J1723--2837. Since there are two extended source within
10$^{\circ}$ of the pulsar (W28 and W30), the {\tt gtdiffrsp} was
first used to compute the diffuse response over this region.

Based on this, we folded only photons with probabilities greater than
0.05 using the Fermi plug-in\footnote{See {\tt
    http://fermi.gsfc.nasa.gov/ssc/data/analysis/user\\/Fermi\_plug\_doc.pdf.}}
for the {\tt tempo2}\footnote{{\tt
    http://sourceforge.net/projects/tempo2/}} pulsar timing package
and the best available radio ephemeris \citep{Craw13}.  The photon
weights were calculated by using the best fit spectral model of the
region around the $\gamma$-ray source.  Folding the extracted
probability-weighted \textit{Fermi} LAT photons with energies $>$100
MeV with the ephemeris of PSR J1723--2837 yields no statistically
significant pulsations. Due to the overwhealming background at low
energies a range of energy bands was also considered, but still no
pulsations were detected.

In principle, at least a portion of the $\gamma$-ray emission
associated with PSR J1723--2837 could arise from the same intra-binary
shock that produces the non-thermal X-ray emission. This scenario has
been proposed by \citet{Tam10} to explain the $\gamma$-ray emission
associated with PSR J1023+0038.  In fact, a similarly steep photon
index to that found in \S7.1 is derived for PSR J1023+0038
($2.9\pm0.2$), suggesting that for PSR J1723--2837 the same
$\gamma$-ray production mechanism dominates. This would make these two
pulsars exceptional among the `redbacks'' in the field of the Galaxy,
which tend to exhibit strong pulsation at the spin period
\citep{Ray12}.  To investigate this possibility, we folded the data at
the binary period but found no statistically significant variability.

\section{CONCLUSION}

We have presented an analysis of \textit{XMM-Newton} and
\textit{Chandra} observations of the nearby PSR J1723--2837
``redback'' MSP system. The X-ray spectrum show a strong non-thermal
component that accounts for most if not all of the emission in the
0.3--8 keV band, which exhibits large-amplitude variability as a
function of the binary orbital period. This pronounced flux
modulation, with a significant decline in flux at conjunction, appears
to be one of the defining characteristics of so-called ``redback''
systems. As such, it can serve as a convenient identifier of
additional members of this population, especially in instances where
radio detection is difficult due to prolonged eclipses. For instance,
strong outflows in similar systems may inhibit detection
of radio pulsations, rendering them perpetually eclipsed
\citep{Tav91}. Such an occurence is believed to be the cause of the recent disappearance of PSR J1023+0038 at radio frequencies \citep{Stap13}.

There is no indication of an X-ray wind nebula associated with the
pulsar.  The lack of a discernable PWN down to a limit of
$\lesssim$$3.6\times10^{29}$ ergs s$^{-1}$, corresponding to
$\lesssim$$7\times10^{-6}$ of the pulsar's $\dot{E}$, indicates that
the combination of the low density of the surrounding interstellar
medium, unfavorable wind geometry, and/or possibly low space velocity
are likely not conducive to the production of an X-ray-bright bow
shock.  Thus, X-ray bow shocks associated with nearby MSPs remain
quite rare, with only two objects, PSRs B1957+21 \citep{Stap03} and
J2124--3358 \citep{Hui06}, exhibiting prominent bow shock emission.

A likelihood analysis of the \textit{Fermi} LAT emission in the
vicinity of PSR J1723--2837 reveals that the pulsar is consistent with
being a $\gamma$-ray point source although owing to the strong diffuse
background a detection cannot be established conclusively. There are
no statistically significant $\gamma$-ray pulsations detected even
using photon probability weights. The absence of pulsed emission could
arise due to one or more of the following reasons: (1) the
$\gamma$-ray emission at the pulsar position is unrelated to the
pulsar indicating that PSR J1723--2837 is sub-luminous in
$\gamma$-rays or its $\gamma$-ray emission pattern is not favorably
oriented; (2) the pulse shape is not favorable (e.g.~due to a high
duty cycle), which combined with the high background and the paucity
of source photons above $300$ MeV results in a non-detection.  There
is also no evidence for orbital-phase-dependent variability if the
$\gamma$-rays.

By analogy with PSR J1824--2452I, the redback and \textit{bona fide}
X-ray binary/radio MSP transition system \citep{Pap13}, PSR
J1723--2837 could also experience a switch to an accretion disk state.
Therefore, as the nearest such system, PSR J1723--2837 warrants close
scrutiny at all wavelengths as it provides the best-suited target for
studying the transition process of MSPs from accretion to rotation
power (and vice versa) and the circumstances surrounding it.

\acknowledgements This work was funded in part by NASA
\textit{Chandra} grants GO2-13049A/B awarded through Columbia University
and West Virginia University and issued by the \textit{Chandra} X-ray
Observatory Center, which is operated by the Smithsonian Astrophysical
Observatory for and on behalf of NASA under contract NAS8-03060.  A
portion of the results presented was based on observations obtained
with \textit{XMM-Newton}, an ESA science mission with instruments and
contributions directly funded by ESA Member States and NASA.  This
research has made use of the NASA Astrophysics Data System (ADS) and
software provided by the Chandra X-ray Center (CXC) in the application
package CIAO. We acknowledge the use of data and software facilities
from the FSSC, managed by the HEASARC at the Goddard Space Flight
Center.

Facilities: \textit{XMM,CXO,Fermi}

\end{document}